\documentclass[english,aps,prl,twocolumn, superscriptaddress,showpacs, amsmath]{revtex4-1}

\usepackage[T1]{fontenc}
\usepackage[latin9]{inputenc}
\usepackage{color}
\usepackage{units}
\usepackage{amssymb}
\usepackage{graphicx}
\usepackage{esint}
\usepackage{bm}
\usepackage{natbib}
\usepackage{ulem}
\usepackage[colorlinks=true, citecolor=red]{hyperref}
\setcitestyle{journalcolor= blue}

\newcommand{\SRO}{SrRuO$_3$ }
\newcommand{\CRO}{CaRuO$_3$ }
\newcommand{\noise}{$\frac{<\delta R^2>}{R^2}$ }
\usepackage{graphicx}
\usepackage{dcolumn}
\usepackage{bm}

\usepackage{babel}

\begin{document}
\title{Probing a spin-glass state in \SRO thin films through higher-order statistics of resistance fluctuations}
\author{Gopi Nath Daptary}
\author{Chanchal Sow}
\author{P. S. Anil Kumar}
\author{Aveek Bid}
\email{aveek.bid@physics.iisc.ernet.in}
\affiliation{Department of Physics, Indian Institute of Science, Bangalore 560012, India}

\begin{abstract}

The complex perovskite oxide \SRO shows intriguing transport properties at low temperatures due to the interplay of spin, charge, and orbital degrees of freedom. One of the open questions in this system is regarding the origin and nature of the low-temperature glassy state. In this paper we report on measurements of higher-order statistics of resistance fluctuations performed in  epitaxial thin films of \SRO  to probe this issue. We observe large low-frequency non-Gaussian resistance fluctuations over a certain temperature range. Our observations are compatible with that of a spin-glass system with properties described by hierarchical dynamics rather than with that of a simple ferromagnet with a large coercivity.

\end{abstract}


\maketitle

\SRO is one the most studied material in condensed matter physics because of its interesting electronic, magnetic, and structural properties~\cite{allen1996transport,okamoto1999correlation,mazin1997electronic,rondinelli2008electronic,klein1996anomalous,jeng2006orbital,kennedy1998high,koster2012structure}. It is the infinite-dimensional member ($ n=\infty $) of the  Ruddlesden-Popper series~\cite{ruddlesden1957new} of ruthenates Sr$_{n+1}$Ru$_n$O$_{3n+1}$, with $n$ denoting the number of Ru-O layers between two alternate layers of Sr-O. The band structure of ferromagnetic \SRO calculated within local spin-density approximation (LSDA)~\cite{singh1996electronic,fujioka1997electronic} shows a strong Ru(4d)-O(2p) hybridization. The resulting high degeneracy of the $t_{2g}$ orbitals ensures that any orbital fluctuation  couples to the spin via spin-orbit coupling~\cite{bradaric2008hidden}, making \SRO an attractive system for studying the interplay of electronic, spin, and orbital degrees of freedom.

Despite intensive research for more than four decades, the low-temperature transport properties of \SRO are far from understood~\cite{koster2012structure}. Recently, it has been shown from magnetic memory effects that there are strong signatures of glassy behavior in bulk samples of \SRO at low temperatures ~\cite{sow2012structural}.  There have been a few reports~\cite{PhysRevB.79.104413,reich1999spin,pi2006exchange,ravindran2002antiferromagnetic} suggesting spin-glass-like behavior in epitaxial thin films of \SRO over a similar temperature range. Measurements show that there is a significant difference between zero-field-cooled and field-cooled magnetization at low temperatures with the emergence of a pronounced cusp in the magnetization at a certain temperature which smoothes out at higher magnetic fields. This has been interpreted~\cite{PhysRevB.79.104413}, in accordance with the ideas of Edwards and Anderson~\cite{0305-4608-5-5-017}, to be due to spin clusters randomly distributed in the matrix; thus  supporting the idea of the existence of a spin-glass state in the system. Theoretical calculations ~\cite{zayak2006structural} predicted that at low temperatures \SRO has A- and C-type antiferromagnetic (AFM) spin configurations co-existing with the dominant ferromagnetic (FM) state. It was conjectured that FM spin clusters at the Ru site may be embedded with some AFM spin clusters at the Sr site, causing the randomness in the system which could be the origin of glassy behavior in this system~\cite{PhysRevB.79.104413}. It has also been suggested that spin fluctuations induced in the system due to coupling of the orbital disorder with magnetic dynamics via spin-orbit coupling leads to a large distribution of  relaxation times and hence glassiness in the system~\cite{bradaric2008hidden}. Alternately,  some reports put down the anomalous low-temperature properties of \SRO to its large coercivity (for a review see Ref. ~\cite{koster2012structure}).

One way to distinguish spin glasses from simple ferromagnetic systems with large coercivity is through the higher-order statistics of resistance fluctuations. It has been shown through a series of experiments ~\cite{WeissmanCuMn,RevModPhys.65.829} that the most probable model for spin glass is the hierarchical kinematic model. In this model, below a certain temperature there exists a large number of possible metastable configurations for a spin-glass system~\cite{RevModPhys.58.801,spinglass}. Consequently, there exists a broad range of characteristic relaxation rates - the transitions between these configurations are what give rise to the slow dynamics in the system. One measurable consequence of this slow dynamics is that the measured power spectral density (PSD) of resistance fluctuations is not static in time - in other words there is significant "spectral wandering." This introduces significant non-Gaussian components (NGC) in the resistance fluctuations. The NGC can be experimentally probed by measuring the higher-order statistics of resistance fluctuations.  On the other hand, in simple ferromagnetic systems with large coercivity such non-Gaussian fluctuations are not expected. To probe the presence of a spin-glass state in \SRO  we have studied the higher-order statistics of low-frequency resistance fluctuations over an extensive temperature and magnetic field range. Our measurements find significant evidence for non-Gaussian fluctuations with characteristics that are consistent with the existence of a spin glass state in \SRO at low temperatures.

\SRO thin films were epitaxially grown on LaAlO$_3$(001) substrate (lattice mismatch of ~$3.4\%$ with \SRO) using pulsed laser deposition (PLD). All the films were grown using KrF ($\lambda$ = 248 nm) laser under the following conditions: (i) 1.6 J/cm$^2$ fluence, (ii) 0.3 mbar oxygen pressure, and (iii) 700 $^\circ$C substrate temperature. After growth the samples were $ in  situ$ annealed at high oxygen pressure for an hour to maintain the right oxygen stoichiometry. We have carried out measurements on six samples - all of them grown under similar conditions and differing only in their physical dimensions. Transport and resistance fluctuation measurements on all the films gave qualitatively similar results. In this paper we discuss in detail the results  from three of   these films of thickness 30 nm. Two of the samples (S1 and S3) are broad films (width 3 mm) whereas the third one (S2) was patterned into a hall bar of width 100 $\mu$m with distance between the two voltage probes 275 $\mu$m.  The structural characterizations were performed using Rigaku out Smart Lab X-Ray diffractometer and it was confirmed that the growth is $c$-axis oriented. Magnetic characterizations were carried out in a Quantum Design Physical Property Measurement system ( PPMS) over the temperature range of 3-300 K in a 0-5 T magnetic field. Resistivity ($\rho$), magnetoresistance (MR) and resistance fluctuation measurements were done over a temperature range of 1.5-300 K in magnetic fields up to 8 T in a Helium-3 cryostat.

\begin{figure}[tbh]
\begin{center}
\includegraphics[width=0.48\textwidth]{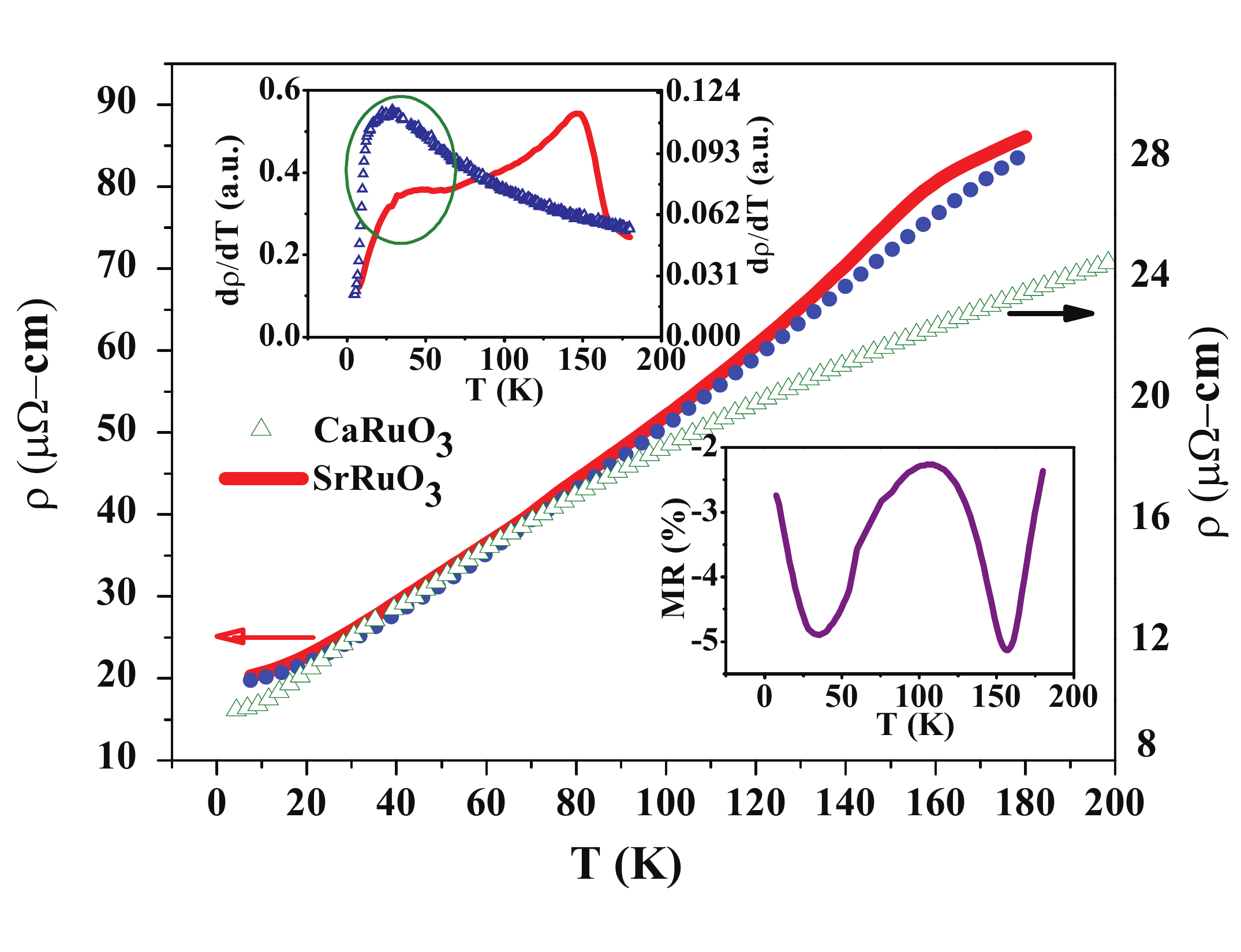}
\small{\caption{ (Color online). (a) Resistivity of \SRO (sample S2) (red solid line) and \CRO (olive open triangles) as a function of temperature at zero magnetic field. Also shown is the resistivity of \SRO at 8 Tesla magnetic field (blue filled circles). Upper inset: Temperature dependence of $d\rho$/$dT$ of \SRO(sample S2) (red solid line) and \CRO (blue open triangles). Lower inset: Magnetoresistance of \SRO as a function of temperature - note the negative MR peaks at $T_c$ and $T^*$. \label{fig:rt}}}
\end{center}
\end{figure}

Figure.\ref{fig:rt} shows the resistivity $\rho$ of sample S2 as a function of temperature at 0 T and at 8 T magnetic fields. Resistivity shows metallic behavior down to 1.5 K with no upturn at low temperatures thus attesting to the high quality of the films. There is a small kink in $\rho$ around the ferromagnetic transition temperature $T_C$. The kink gets suppressed under an 8 T magnetic field applied  perpendicular to the plane of the film. This is a feature commonly observed during a transition from paramagnetic to a ferromagnetic state due to the rapid decrease of scattering from spin disorder and can be well explained by the Fisher Langer theory~\cite{PhysRevLett.20.665}. Below 20 K the resistivity has a quadratic dependence on the temperature expected for a Fermi liquid (FL). The coefficient of the quadratic term does not  change significantly in the presence of an 8 T magnetic field, showing that the resistivity arises due to electron-electron scattering rather than electron-magnon scattering which also has a quadratic dependence on the temperature.  A careful inspection of the plot of $d\rho/dT$ as a function of temperature reveals a broad hump at around 37 K (the region marked by the circle  in the upper inset of Fig. \ref{fig:rt}), we denote this characteristic temperature as $T^*$. The magnetoresistance also shows large negative peak around this temperature (see Fig. \ref{fig:rt} lower inset). This low-temperature anomaly in the resistivity and magnetoresistance is also seen in polycrystalline and single crystal samples. Interestingly, the coercive field measured in bulk samples has a maxima at around the same temperature~\cite {hou2002zero,sow2012structural}.

\begin{figure}[tbh]
\begin{center}
\includegraphics[width=0.48\textwidth]{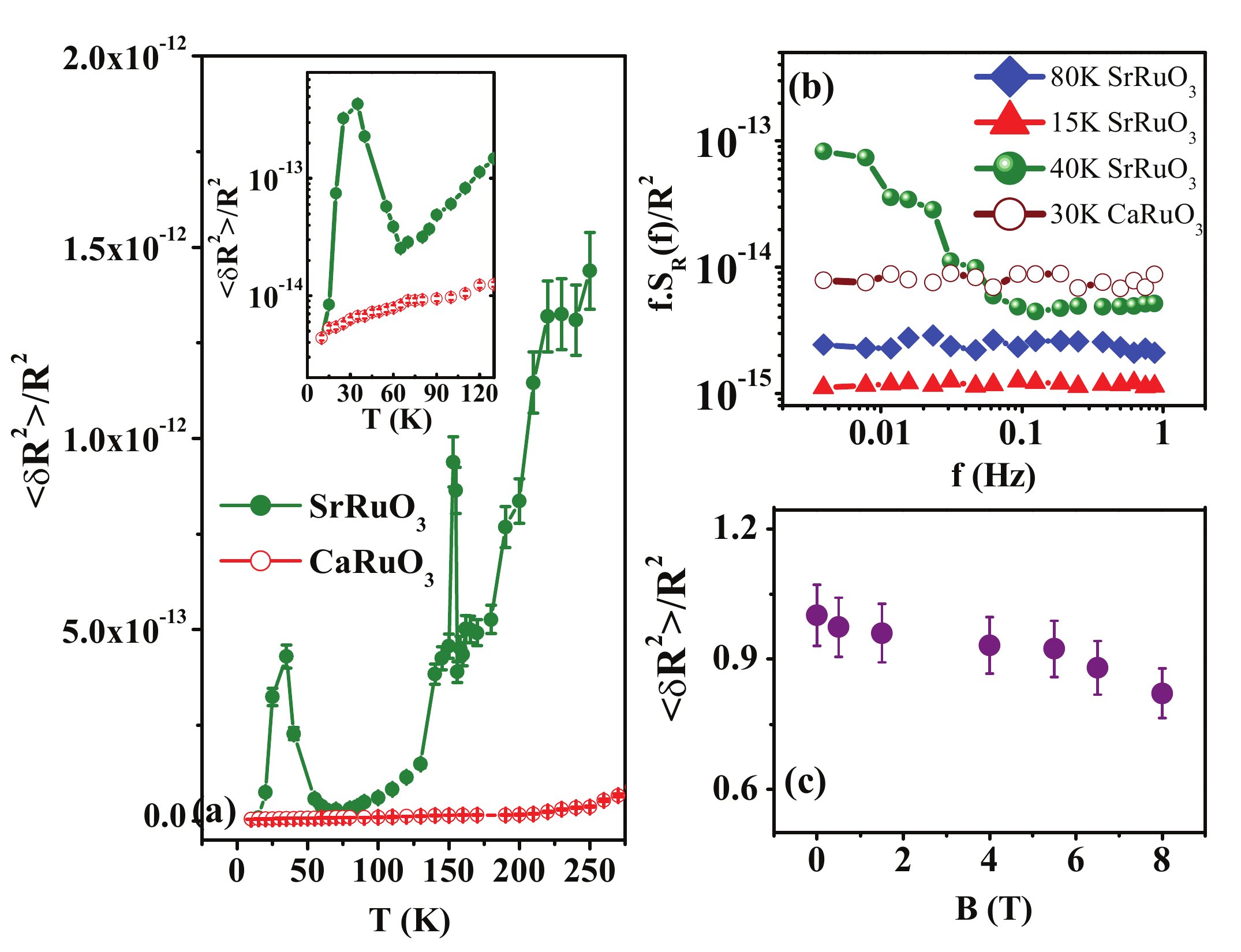}
\small{\caption{(Color online). (a) Relative variance of resistance fluctuations \noise as a function of temperature of \SRO (olive filled circles) and \CRO (red open circles), respectively. In the case of \SRO note the peaks in \noise  at $T=T_c$ and at $T=T^*$. The inset shows \noise in the low-temperature range on a linear-log scale to emphasize its sharp decrease in the FL region (see text for details). (b) Scaled PSD, $f . S_R(f)/R^2$ at a few representative temperatures for \SRO.  The data for \CRO (brown open circles) are also plotted for comparison. The data at different temperatures have been shifted vertically for clarity. (c) Relative variance of resistance fluctuations  at 25 K as a function of magnetic field for \SRO. The data have been normalized with the 0 T value. \label{fig:noise}}}
\end{center}
\end{figure}

\begin{figure}[tbh]
\begin{center}
\includegraphics[width=0.48\textwidth]{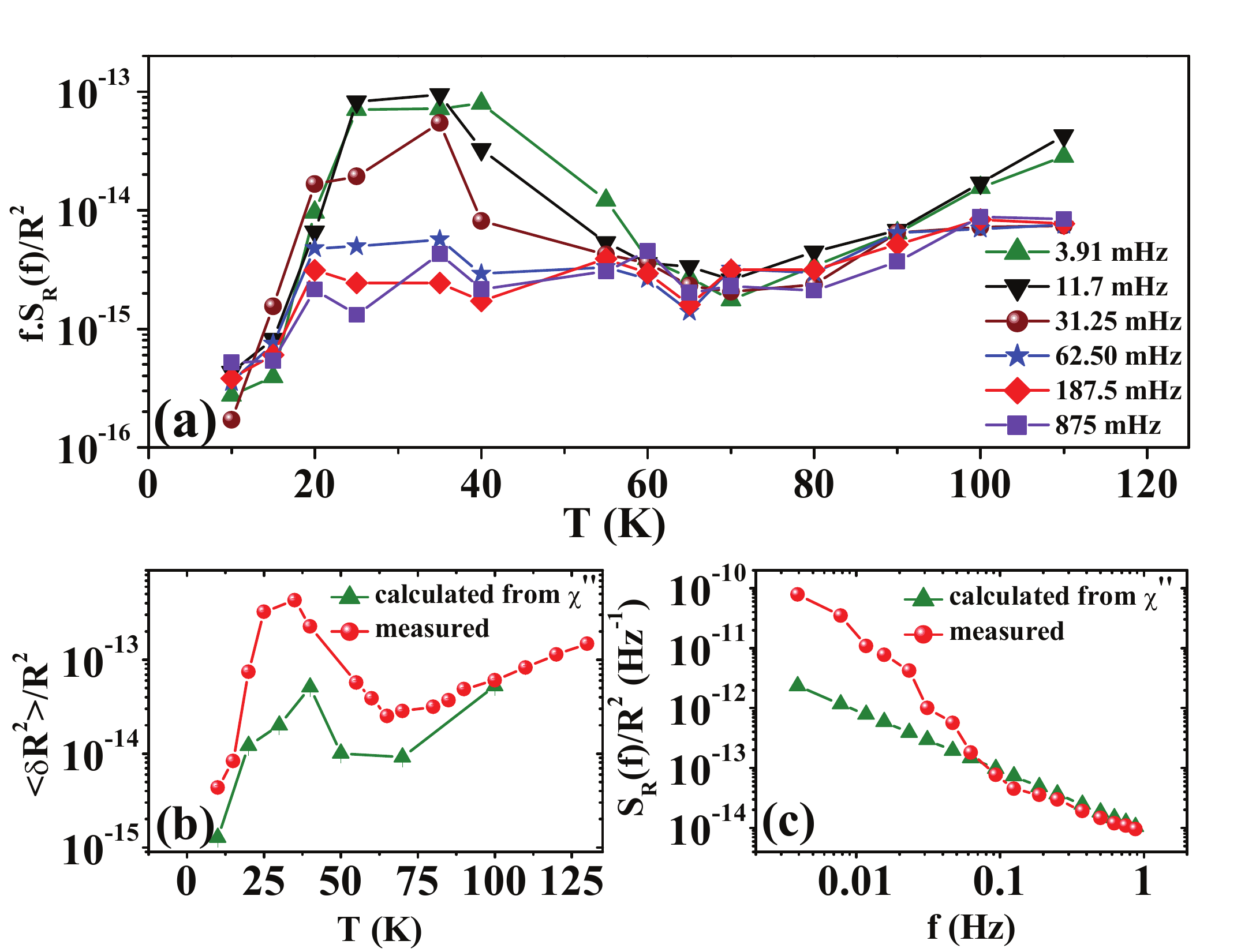}
\small{\caption{(Color online). (a) Plot of scaled PSD at different frequencies as a function of temperature. (b) Temperature dependence of \noise calculated from $\chi^{''}$ using FDT (olive triangles). For comparison we also plot the measured values of \noise (red circles). (c) Comparison of the PSD measured  at 40 K (red circles) and the PSD calculated using FDT (olive triangles) from $\chi^{''}$ at the same temperature. \label{fig:noise_chi}}}
\end{center}
\end{figure}

To probe further the nature of transport at low temperatures we  measured low-frequency resistance fluctuations in samples S1 and S2 using standard ac 5-probe technique~\cite{scofield1987ac}. The power spectral density (PSD) of resistance fluctuations $S_R(f)$ was calculated from the time series of resistance fluctuations accumulated using a fast analog to digital convertor ( ADC)  card ~\cite{ghosh2004set}. The PSD was integrated over the bandwidth of measurement to get the relative variance of resistance fluctuations \noise at a given temperature. In Fig. \ref{fig:noise}(a) we show a plot of \noise measured for sample S1 as a function of temperature. The data from sample S2 was qualitatively similar after taking into account the scaling of noise with volume. The fact that the noise scaled inversely to the volume (or equivalently the number of carriers) confirms that the low-frequency noise in these systems arises from fluctuations in the sample mobility and \emph{not} the number density. We note that, in addition to the expected peak in noise at $T_c$ ~\cite{reutler2000local} there is also a broad peak in noise centered around $T^*$. On further cooling the sample down to 1.5 K, the resistance fluctuations decrease by almost two orders of magnitude. In Fig. \ref{fig:noise}(b) we plot the PSD as a function of frequency at a few representative temperatures. The data are plotted as $fS_R(f)/R^2$ to accentuate the deviation of the PSD from the $1/f$ dependence. It can be seen that at all temperatures both above and below $T^*$ the PSD is of the type $S_R(f) \propto 1/f^\alpha$ with the value of $\alpha \sim 1$. Near $T^*$ the spectrum deviated significantly from $1/f$ nature in the low-frequency region. This can be seen more clearly from the plots in Fig.\ref{fig:noise_chi}(a) where $fS_R(f)/R^2$ at different frequencies is plotted as a function of temperature. At temperatures away from $T^*$, all the plots (corresponding to fluctuations at different frequencies) collapse on top of each other attesting to the $1/f$ nature of the fluctuations. At temperatures near $T^*$, the scaled low-frequency noise deviates significantly from the high-frequency components. This seems to indicate  that there are at least two distinct processes giving rise to resistance fluctuations in the system. One process induces resistance fluctuations which are essentially $1/f$ in nature and is present over the entire temperature range. The second process produces an excess low-frequency resistance fluctuation solely near $T^*$.

A major component of the $1/f$ noise in spin-glass systems arises due to magnetization fluctuations through the fluctuation-dissipation theorem (FDT)
\begin{equation}
S_R^M(f)= \frac{2V}{\pi}\frac{k_BT\chi^{''}}{f}(\frac{\partial R}{\partial M})^2 ,
\label{eqn:fdt}
\end{equation}
where $\chi^{''}$ is the imaginary part of the magnetic susceptibility, $V$ is the sample volume, and $M$ is the magnetization of the film~\cite{kogan2008electronic}. The accurate measurement of $\chi^{''}$ in thin films is extremely challenging and the data obtained are often not reliable. We have instead measured $\chi^{''}$ in bulk samples of \SRO and used the data to get an estimate of the temperature dependence of $\chi^{''}$ in  our thin films. Using this value of $\chi^{''}$, the temperature dependence of the relative variance of $S_R^M(f)$ was calculated for sample S3 from Eq.~ (\ref{eqn:fdt}) and is shown in Fig. ~\ref{fig:noise_chi}(b). It follows a pattern very similar to that of $S_R(f)$ for sample S1 with a sharp peak near $T^*$ establishing that a significant part of the fluctuations arise from the FDT.  The detailed spectra of $S_R(f)$ and $S_R^M(f)$, however, show important differences - the spectra for a particular temperature ($T/T*=1.1$) is shown in Fig.~\ref{fig:noise_chi}(c). As seen from Eq. (\ref{eqn:fdt}), $S_R^M(f)$ has a $1/f$-type spectra over the entire frequency range of interest. For frequencies larger than 50 mHz,  $S_R(f)$ has a $1/f$ dependence on the frequency and its value closely matches the estimated value for $S_R^M(f)$. At lower frequencies, however, $S_R(f)$ deviates significantly from $S_R^M(f)$ showing that there are additional slow dynamics in the system that can not be accounted for solely by the FDT as in a canonical spin-glass state.

Motivated by the observation of negative magnetoresistance peak at $T=T^*$, we have studied the resistance fluctuations at different temperatures near $T=T^*$ in the presence of a magnetic field. Measurements were done in magnetic fields up to 8 T, which is larger than the coercive field in these films at these temperatures. Figure.\ref{fig:noise}(c)shows the dependence of the relative variance of resistance fluctuations at $T \simeq 25$ K as a function of magnetic field measured in sample S2, the data have been scaled to the value at 0 T. We note that the noise  is only partially suppressed by an 8 T magnetic field. As we show in the discussion section, this is compatible with what is expected for a canonical spin glass system~\cite{PhysRevLett.63.794}. It should also be noted that \SRO possesses high uniaxial magneto-crystalline anisotropy (anisotropy field approximately 10 T) due to the spin-orbit coupling of the Ru atoms.~\cite{klein2000domain}). We also have found that the Arrot plots (for determing critical exponents) in the critical region become parallel straight lines only above 8 T.

\begin{figure}[tbh]
\begin{center}
\includegraphics[width=0.48\textwidth]{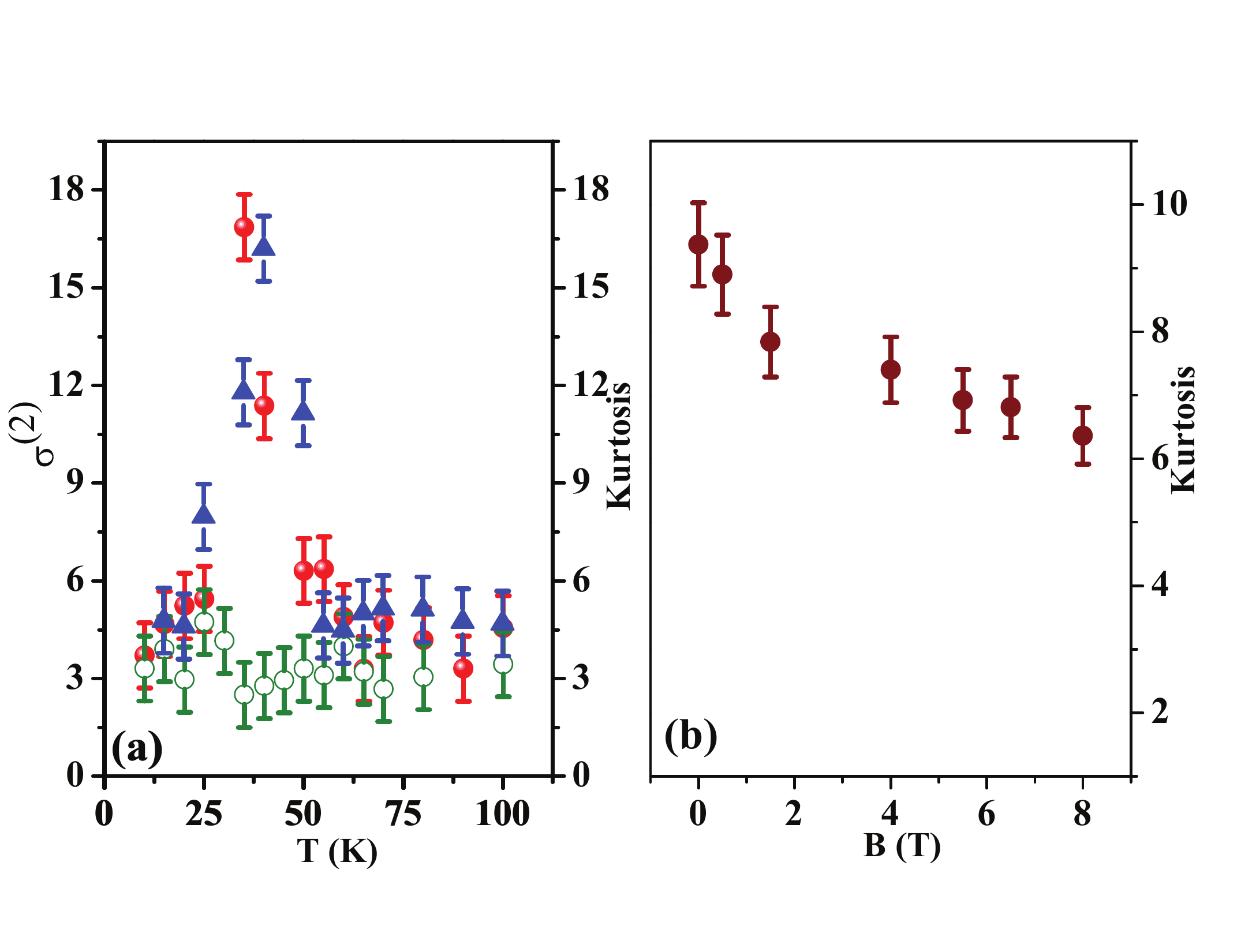}
\small{{\caption{(Color online). (a) Plot of normalized second spectrum $\sigma^{(2)}$ of resistance fluctuations measured for \SRO film (red filled circles) and for \CRO film (olive circles). Also plotted is the kurtosis of \SRO (blue triangles) as function of temperature at 0 T magnetic field. (b) Plot of the kurtosis of resistane fluctuations in \SRO film as function of magnetic field measured at 25 K. \label{fig:2spec}}}}
\end{center}
\end{figure}

Higher-order statistics of resistance fluctuations have been used extensively to study the presence of long range correlations in systems undergoing electronic, magnetic, or spin-glass transitions~\cite{koushik2013correlated,chandni2009criticality,restle1985non,seidler1996dynamical}. If the fluctuators in a system are independent of each other, then the central limit theorem~\cite{reif2009fundamentals} guarantees that the fluctuation statistics would be Gaussian. However, if there is the presence of long-range correlations in the system due to magnetic, electronic, or structural interactions, then the resultant time-dependent fluctuation statistics have a strong non-Gaussian component. To probe specifically the existence of a spin-glass state in this system at low temperatures we calculated higher order statistics of resistance fluctuations using two different methods: (1) second spectrum and (2) kurtosis. The second spectrum is the Fourier transform of the four-point correlations of resistance fluctuations filtered over a chosen frequency octave ($f_l$, $f_h$). It is mathematically defined as
\begin{equation}
S_R^{f_1}(f_2)=\int_0^\infty \langle\delta R^2(t)\rangle\langle\delta R^2(t+\tau)\rangle cos(2\pi f_2\tau)d\tau ,
\end{equation}
where $f_1$ is the center frequency of a chosen octave and $f_2$ is the spectral frequency. Details of the calculation method are given elsewhere ~\cite{koushik2013correlated}. Physically, $S_R^{f_1}(f_2)$ represents the "spectral wandering" or fluctuations in the PSD with time in the chosen frequency octave. To avoid corruption of the signal by the Gaussian background noise we have calculated the second spectrum over the frequency octave 31.25-62.5 mHz, where the sample noise is significantly higher than the background noise. A convenient way of representing the second spectrum is through the normalized second spectrum $\sigma^{(2)}$ defined as
\begin{equation}
\sigma^{(2)}=\int_0^{f_h-f_l}S_R^{f_1}(f_2)df_2/[\int_{f_l}^{f_h}S_R(f)df]^2 ,
\end{equation}
Figure. \ref{fig:2spec}(a) shows $\sigma^{(2)}$ as a function of $T$ at zero magnetic field - the values have been suitably normalized with the Gaussian background~\cite{RevModPhys.65.829}. $\sigma^{(2)}$ peaks near $T=T^*$ and decays to the Gaussian background value of 3 to either side of it. The large increase of $\sigma^{(2)}$ near $T=T^*$ suggests that there is a rapid build up of correlations in the system as this characteristic temperature is approached from either side. To verify this observation using the time-dependent fluctuation data directly we computed their kurtosis ~\cite{decarlo1997meaning,mansour1998kurtosis}. Kurtosis can be defined as the normalized fourth moment of a time series: $\beta=\mu_4/\sigma^2$; $\mu_4$ is the fourth moment about the mean and $\sigma$ the second moment. For any random process which follows a Gaussian distribution, the kurtosis $\beta$ equals 3. Any deviation from 3 is indicative of a non-Gaussian distribution. In particular $\beta >3$, i.e., positive excess kurtosis suggests that the system has long tails in the distribution function indicating large cooperative jumps. Kurtosis for sample S1 calculated from the time series of voltage fluctuations at zero magnetic field in the frequency ranges 93.75-187.5 mHz is shown in Fig. \ref{fig:2spec}(a). The NGC peaks near $T=T^*$ , while decreasing to the baseline value on either side of it, exactly following the behavior of $\sigma^{(2)}$.

To understand whether these observations are peculiar to \SRO we have carried out similar measurements on \CRO thin films grown under similar conditions. \CRO has a very similar structure and lattice constant to \SRO but it does not develop any magnetic order over any temperature range. The resistivity measurements on \CRO show a similar hump in the $d\rho/dT$. But unlike \SRO , there is no distinctive noise signature in this temperature region - the resistance fluction is Gaussian and has a $1/f$ nature over the entire temperature range spanning 5 to 300 K. This strongly suggests that the non-gaussianity and long time scales observed in the resistance fluctuations in \SRO devices is not merely of structural origin, but arise most probably from the underlying magnetic nature of the ground state.

To summarize our observations on \SRO: near $T^*$ we have  (i) a hump in $d\rho/dT$ , (ii) a small negative peak in magnetoresistance, (iii) a large increase in resistance fluctuations, and (iv) a large non-Gaussian component in the resistance fluctuation spectra. Our observations are consistent with the idea that in the temperature range between around 20-70 K there exists a spin-glass state in the system. The glassiness probably arises due to the extremely slow relaxation of the magnetization seen in \SRO around these temperatures,  which in turn is due to large domain misfits~\cite{PhysRevB.79.104413,sow2012structural}.  As shown in Fig.~\ref{fig:noise_chi}, the large noise seen in this temperature range has two distinct components. One component has a $1/f$ spectrum over the entire temperature range and its magnitude matches quite well with what is expected from the FDT for a spin-glass system. In addition, there is a second component which peaks at low frequencies and causes the spectrum to deviate from the $1/f$ nature in this region. The origin of this excess low-frequency noise is not clear at present and understanding it needs further studies. A probable source may be the coupling of orbital fluctuations to spin fluctuations through spin-orbit interactions. Recent studies ~\cite{PhysRevLett.100.096402,bradaric2008hidden} have shown that in \SRO near $T=T^*$ there is significant scattering of the charge carriers due to orbital fluctuations. Such charge scattering by orbital fluctuations are known to produce  excess low-frequency resistance noise in the system~\cite{bid2003low}.

The large non-Gaussianity in the resistance fluctuations strongly indicates the presence of a spin-glass state with hierarchical kinetics in \SRO in the temperature range around $T^*$. For spin-glass systems the integrated noise power as well as the non-Gaussianity in the fluctuation spectrum do not get affected significantly by a moderate magnetic field.  This is consistent with our observation that the excess non-Gaussian noise is only partially suppressed under an 8 T magnetic field [see Fig.\ref{fig:2spec}(b)]. Eventually at very low temperatures the transport properties of \SRO begin to resemble that of a Fermi liquid - the noise spectra regains it $1/f$ nature and the integrated noise decreases rapidly with decrease in temperature.

To conclude, we have studied in detail higher- order statistics of resistance fluctuations in thin films of \SRO in the low- temperature region where it displays anomalous transport properties.  We observe large excess non-Gaussian resistance fluctuations - the characteristics of the fluctuations are consistent with that of a spin- glass system with properties described by hierarchical dynamics rather than that of a simple ferromagnet with a large coercivity.

\begin{acknowledgments}
We acknowledge funding from Nanomission,  Department of Science \& Technology (DST) and the Indian Institute of Science (IISc), Bangalore.
\end{acknowledgments}

\bibliographystyle{apsrev4-1}
\bibliography{sro}

\end{document}